\documentclass[a4paper,10pt]{article}
\usepackage[cp1251]{inputenc}
\usepackage[english]{babel}
\usepackage{amsmath}
\usepackage{amssymb}
\usepackage{graphicx}
\usepackage[pdftex]{color}
\usepackage[square,numbers,comma,sort&compress]{natbib}
\usepackage[pdftex]{hyperref}
\hypersetup{
             final,
             colorlinks=true,
             linkcolor=blue,
             citecolor=red
}

\DeclareMathOperator{\Tr}{Tr}

\DeclareMathOperator{\sign}{sgn}

\begin{document}
\twocolumn[\fontsize{6pt}{7pt}\selectfont\textsl{ISSN 0021-3640, JETP Letters, 2016, Vol. 103, No. 12, pp. 774--779. \textcopyright\,Pleiades Publishing, Inc., 2016.\\ Original Russian Text \textcopyright\,P.\,V. Ratnikov, 2016, published in Pis'ma v Zhurnal Eksperimental'noi i Teoreticheskoi Fiziki, 2016, Vol. 103, No. 12, pp. 872--877.}

\vspace{0.18cm}

\begin{center}
\rule{7cm}{0.4pt}\hspace{3cm}\rule{7cm}{0.4pt}

\vspace{-0.15cm}
\rule{7cm}{0.4pt}\hspace{3cm}\rule{7cm}{0.4pt}

\normalsize
\vspace{-0.55cm}
{\bf CONDENSED\\
MATTER}

\vspace{0.5cm}

\LARGE{\bf On the Dispersion Relation of Magnetoplasmons\\
in a Planar Graphene-Based Superlattice}

\vspace{0.5cm}

\large{\bf P.\,V. Ratnikov}

\vspace{0.15cm}

\normalsize

\textit{Prokhorov General Physics Institute, Russian Academy of Sciences, ul. Vavilova 38, Moscow, 117942 Russia\\
e-mail: ratnikov@lpi.ru}

\vspace{0.25cm}

Received May 4, 2016; in final form, May 10, 2016
\end{center}

\vspace{0.1cm}
\begin{list}{}
{\rightmargin=1cm\leftmargin=1cm}
\item
\small{The dispersion relation for magnetoplasmons in a planar superlattice with periodically alternating regions of gapless and gapped modifications of graphene has been derived within the frame of the random-phase approximation. The contribution of virtual transitions between the lower electron miniband and the upper hole miniband to the polarization operator has been taken into account in addition to the contribution of virtual intra-miniband transitions.}

\vspace{0.05cm}

\normalsize{\bf DOI}: 10.1134/S0021364016120092

\end{list}\vspace{0.35cm}]

\begin{center}
1. INTRODUCTION
\end{center}

Collective excitations in gapless graphene in a magnetic
field were studied theoretically in a number of
works \citep{Lozovik, Berman}. However, less attention was paid to collective
excitations in a gapped modification of
graphene in a magnetic field.

In this work, magnetoplasmons in a planar superlattice
based on gapless graphene and its gapped modification
are considered (\hyperlink{fig1}{Fig. 1}). We use a model that
allows finding analytically the dispersion relation for
charge carriers in such a superlattice \citep{Ratnikov1}. The dispersion
relation of plasmons in this superlattice was found
in \citep{Ratnikov2}.

The extrema of the bands of both the gapped and
gapless modifications of graphene in the \textbf{k} space lie at
the $K$ and $K^\prime$ points of the Brillouin zone. The band
structure of the gapped modifications of graphene also
contains two valleys. The basic difference from gapless
graphene is the presence of the energy gap between the
extrema of the conduction and valence bands. In the
general case, the center of the gap is displaced in
energy from the position of the $K$ and $K^\prime$ points of gapless
graphene (the level $E = 0$) by the work function.

Alternating strips of gapless and gapped graphene
provide modulation in the space of the energy gap.
This is equivalent to the application of a one-dimensio-nal
periodic potential. The energy spectrum of the
systems splits into smaller bands called minibands separated
by minigaps.

In this work, we suggest a model for the description
of the superlattice under consideration, in which the
charge carriers in different valleys interact identically
with the magnetic field. In the general case, such a valley
degeneracy can be broken, e.g., by a uniaxial strain in
the graphene plane \cite{Bir}.

\newpage
\begin{center}
2. WAVEFUNCTIONS\\
AND THE SINGLE-PARTICLE ENERGY\\
SPECTRUM OF CHARGE CARRIERS
\end{center}

In our previous work \citep{Ratnikov2}, we introduced the effective
Hamiltonian of charge carriers in the valley of the
$K$ point of a graphene-based superlattice in zero magnetic
field. Two case were considered: (i) a quasi-one-dimensional
case (the Fermi level falls within a minigap)
and (ii) a quasi-two-dimensional case (the Fermi level
is located within the miniband).

\begin{figure}[b!]
\begin{center}
\hypertarget{fig1}{}
\includegraphics[width=0.5\textwidth]{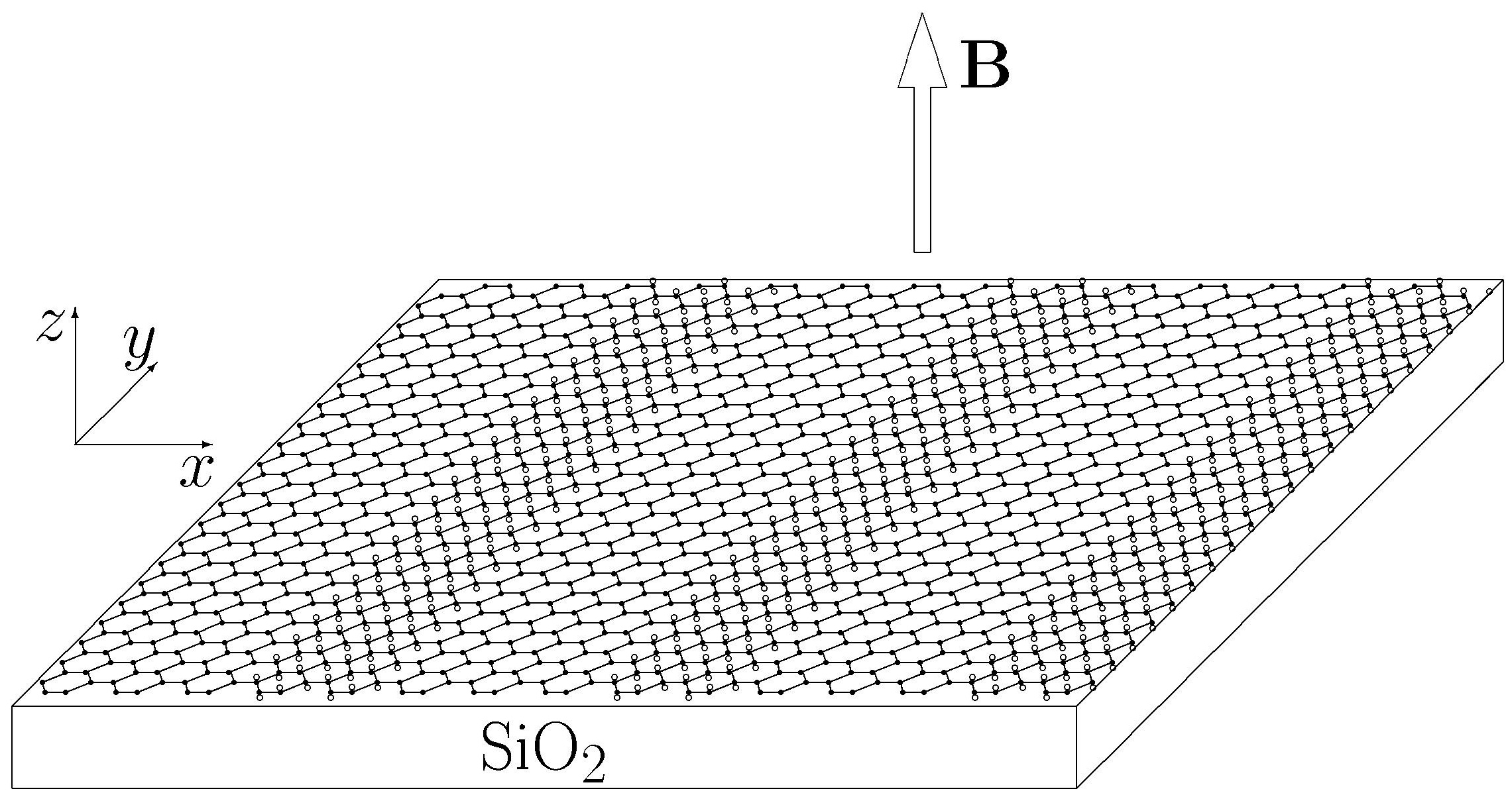}
\end{center}

\begin{list}{}
{\rightmargin=0.27cm\leftmargin=0.27cm}
\item
\footnotesize{\bf Fig. 1.} Example of a system under consideration: a
graphene-graphane superlattice on a SiO$_2$ substrate in the
magnetic field \textbf{B} (the positions of hydrogen atoms are
shown by open circles).
\end{list}
\normalsize
\end{figure}

In this work, we will assume that a more general
quasi-two-dimensional case takes place in zero magnetic
field. In the presence of the magnetic field, we
replace the momentum operator $\widehat{\bf p}=-i{\boldsymbol\nabla}$ (in units of
$\hbar=1$) in the effective Hamiltonian of charge carriers
in the graphene superlattice by the operator $\widehat{\boldsymbol\pi}=\widehat{\bf p}-\frac{e}{c}{\bf A}$, where $e$ is the charge of the particle, $c$ is the speed of light in vacuum, and ${\bf A}=(0,\,Bx,\,0)$ is the vector potential of the magnetic field ${\bf B}=(0,\,0,\,B)$. The
$(x,\,y)$ plane coincides with the superlattice plane (see
\hyperlink{fig1}{Fig.~1}). In this case, the effective Hamiltonian of
charge carriers in the valley of the $K$ point has the form
\begin{equation}\label{1}
\widehat{H}^K_\text{eff}=\texttt{v}_\perp\sigma_x\widehat{\pi}_x+\texttt{v}_\parallel\sigma_y\widehat{\pi}_y-\sigma_z\Delta_\text{eff}+V_\text{eff},
\end{equation}
where $\texttt{v}_\perp$ and $\texttt{v}_\parallel$ are the effective Fermi velocities
across and along the strips, respectively, of gapless and
gapped graphene and, additionally, $\texttt{v}_\parallel\approx\texttt{v}_\text{F}$ ($\texttt{v}_\text{F}$ is the Fermi velocity in gapless graphene), whereas $\texttt{v}_\perp\ll\texttt{v}_\parallel$
owing to a low probability of tunneling of charge carriers
through the regions of gapped graphene. The
velocities $\texttt{v}_\perp$ and $\texttt{v}_\parallel$ were chosen such that the dispersion
relation of charge carriers in zero magnetic field
would agree with the dispersion relation for the superlattice~\citep{Ratnikov2}.

The Pauli matrices $\sigma_x,\,\sigma_y$, and $\sigma_z$ act in the space of
two sublattices of the hexagonal lattice of graphene
and its gapped modification.

For simplicity, we consider here the lower electron
and upper hole minibands. They are separated by the
minigap $2\Delta_\text{eff}$. If the work function of the gapped modifications
is nonzero, the center of this minigap is displaced
in energy with respect to the $K$ and $K^\prime$ points of
gapless graphene by the effective work function $V_\text{eff}$.

Assuming that $|V_\text{eff}|<\Delta_\text{eff}\ll\Delta_0$ ($\Delta_0$ and $V_0$ are the
half-width of the band gap and the work function of
the gapped modification of graphene, respectively),
we find
\begin{equation}\label{2}
\begin{split}
\Delta_\text{eff}&=\frac{\pi\texttt{v}_\text{F}}{2d_\text{I}}\left[1-\frac{\texttt{v}_\text{F}}{d_\text{I}\Delta_0}\right],\\
V_\text{eff}&=\frac{\texttt{v}_\text{F}}{d_\text{I}\Delta_0}V_0,
\end{split}
\end{equation}
where $d_\text{I}$ is the width of gapless graphene strips with
$\texttt{v}_\text{F}/d_\text{I}\ll\Delta_0$. The additional condition of applicability
of Eqs. \eqref{2} is the inequality $\Delta_0\gtrsim2\texttt{v}_\text{F}/d_\text{II}$, where $d_\text{II}$ is the width of gapped graphene strips \citep{Ratnikov1}.

The effective Hamiltonian of charge carriers in the
valley of the $K^\prime$ point is
\begin{equation}\label{3}
\widehat{H}^{K^\prime}_\text{eff}=\texttt{v}_\perp\sigma_x\widehat{\pi}_x-\texttt{v}_\parallel\sigma_y\widehat{\pi}_y+\sigma_z\Delta_\text{eff}+V_\text{eff}.
\end{equation}
The mass term (the third one) is written with the
opposite sign compared to Hamiltonian \eqref{1}. This provides
the unitary equivalence of $\widehat{H}^K_\text{eff}$ and $\widehat{H}^{K^\prime}_\text{eff}$. They
can be transformed to one another by the unitary transformation
\begin{equation}\label{4}
\widehat{H}^{K^\prime}_\text{eff}=U\widehat{H}^K_\text{eff}U^\dagger,\hspace{0.1cm}U=\sigma_x.
\end{equation}
The immediate consequence of this fact is the equivalence
of their energy spectra.

Let us introduce the dimensionless variable
\begin{equation}\label{5}
\xi=\sqrt{\frac{\texttt{v}_\parallel}{\texttt{v}_\perp}}\left(\frac{x}{l_B}+l_Bk_y\right),
\end{equation}
where $k_y$ is the $y$ component of the crystal momentum and
\begin{equation}\label{6}
l_B=\sqrt{\frac{c}{|e|B}}
\end{equation}
is the magnetic length. The wavefunctions of Hamiltonian \eqref{1}
for the Landau levels $n=1,\,2,\,\ldots$ of electrons and holes are
\begin{equation}\label{7}
\begin{split}
\Psi^{Ke(\pm)}_{nk_y}(x,\,y)&=C^{(\pm)}_n\begin{pmatrix}A^{(\pm)}_n\phi_{n-1}(\xi)\\ \phi_n(\xi)\end{pmatrix}\frac{e^{ik_yy}}{\sqrt{L_y}},\\
\Psi^{Kh(\pm)}_{nk_y}(x,\,y)&=C^{(\pm)}_n\begin{pmatrix}\phi_n(\xi)\\ -A^{(\pm)}_n\phi_{n-1}(\xi)\end{pmatrix}\frac{e^{-ik_yy}}{\sqrt{L_y}},
\end{split}
\end{equation}
where $L_y$ is the size of the system along the $y$ axis and
the +(--) sign corresponds to positive (negative)-frequency
solutions. Here, for definiteness, we regard the
positive-frequency solutions to be the solutions corresponding
to the signs of the particle energy used in this
work, namely, $E>0$ for electrons and $E<0$ for holes;
the negative-frequency solutions can be found from
the positive-frequency ones by changing the sign of
the energy $\widetilde{E}=E-V_\text{eff}$. In Eq. \eqref{7}, we introduced the
coefficients
\begin{equation*}
A^{(\pm)}_n=\mp i\frac{\sqrt{2\texttt{v}_\perp\texttt{v}_\parallel n}}{l_B(\varepsilon_n\pm\Delta_\text{eff})},\hspace{0.1cm}
C^{(\pm)}_n=\left(1+|A^{(\pm)}_n|^2\right)^{-1/2},
\end{equation*}
where $\varepsilon_n=\sqrt{\Delta^2_\text{eff}+2\texttt{v}_\perp\texttt{v}_\parallel\frac{|e|}{c}Bn}$, and functions
\begin{equation*}
\phi_n(\xi)=\frac{H_n(\xi)e^{-\xi^2/2}}{\left(2^nn!\pi^{1/2}l^*_B\right)^{1/2}},
\end{equation*}
where $l^*_B=\sqrt{\frac{\texttt{v}_\perp}{\texttt{v}_\parallel}}l_B$ and $H_n(\xi)$ are the Hermite polynomials.

The positive-frequency solutions for the Landau level $n = 0$,
\begin{equation}\label{8}
\begin{split}
\Psi^{Ke(+)}_{0k_y}(x,\,y)&=\begin{pmatrix}0\\ \phi_0(\xi)\end{pmatrix}\frac{e^{ik_yy}}{\sqrt{L_y}},\\
\Psi^{Kh(+)}_{0k_y}(x,\,y)&=\begin{pmatrix}\phi_0(\xi)\\ 0\end{pmatrix}\frac{e^{-ik_yy}}{\sqrt{L_y}}
\end{split}
\end{equation}
coincide with solution \eqref{7} for $n~=~0$ if one takes into
account that $A^{(+)}_0=0$ and $C^{(+)}_0=1$. However, the situation
with the negative-frequency solutions for $n~=~0$ is different: an uncertainty of the form 0/0 emerges in
the coefficient $A^{(-)}_0$. These solutions are
\begin{equation}\label{9}
\begin{split}
\Psi^{Ke(-)}_{0k_y}(x,\,y)&=C^{(-)}_0\begin{pmatrix}\widetilde{\phi}_0(\xi)\\ \phi_0(\xi)\end{pmatrix}\frac{e^{ik_yy}}{\sqrt{L_y}},\\
\Psi^{Kh(-)}_{0k_y}(x,\,y)&=C^{(-)}_0\begin{pmatrix}\phi_0(\xi)\\-\widetilde{\phi}_0(\xi)\end{pmatrix}\frac{e^{-ik_yy}}{\sqrt{L_y}},
\end{split}
\end{equation}
where
\begin{equation*}
\begin{split}
\widetilde{\phi}_0(\xi)&=\frac{i\pi^{1/4}l^{*1/2}_B\Delta_\text{eff}}{\texttt{v}_\perp}\left[1-\Phi(\xi)\right]e^{\xi^2/2},\\
\Phi(\xi)&=\frac{1}{\sqrt{\pi}}\int\limits_0^{\xi^2}\frac{e^{-t}}{\sqrt{t}}dt.
\end{split}
\end{equation*}
The normalization factor $C^{(-)}_0$ is

\begin{equation*}
\begin{split}
C^{(-)}_0&=(1+I)^{-1/2},\hspace{0.1cm}I=\int\limits_{-\infty}^\infty|\widetilde{\phi}_0|^2dx=\frac{\sqrt{\pi}l^{*2}_B\Delta^2_\text{eff}}{\texttt{v}^2_\perp}\widetilde{I},\\
\widetilde{I}&=\int\limits_{-\infty}^\infty e^{\xi^2}\left[\Phi(\xi)-1\right]^2d\xi\approx0.78.
\end{split}
\end{equation*}

The wavefunctions of Hamiltonian \eqref{3} are found
from wavefunctions of Hamiltonian \eqref{1} by transformation \eqref{4}:
\begin{equation}\label{10}
\Psi^{K^\prime e,h(\pm)}_{nk_y}(x,\,y)=\sigma_x\Psi^{Ke,h(\pm)}_{nk_y}(x,\,y).
\end{equation}
It is worth mentioning that the wavefunctions of electrons
and holes in both valleys are coupled by the charge conjugation transformation\footnote{In relativistic quantum theory, the charge conjugation operator is
defined as $\widehat{\mathcal{C}}=i\gamma_2\widehat{\mathbb{C}}$, where $\gamma_2=\gamma_0\alpha_2$ is the $\gamma$ matrix expressed in terms of the $\alpha$ matrix involved in the Dirac Hamiltonian in
the term including $\widehat{p}_y$; in the standard representation, $\gamma_0=\beta$ is
the matrix involved in the mass term \citep{Bjorken}. Here, $\alpha_2=\sigma_y$ and $\beta=\sigma_z$; hence, $\gamma_2=-i\sigma_x$.}
\begin{equation}\label{11}
\Psi^{K,K^\prime h(\pm)}_{nk_y}(x,\,y)=\widehat{\mathcal{C}}\Psi^{K,K^\prime e(\pm)}_{nk_y}(x,\,y),
\end{equation}
where $\widehat{\mathcal{C}}=\sigma_x\widehat{\mathbb{C}}$ is the charge conjugation operator and $\widehat{\mathbb{C}}$ is the complex conjugation operator.

The energy spectrum of Hamiltonians~\eqref{1}~and~\eqref{3}~is
\begin{equation}\label{12}
E^{e,h}_n=V_\text{eff}\pm\varepsilon_n,
\end{equation}
where the plus (minus) sign stands for electrons (holes).

It should be emphasized that the description of the
system under consideration by effective Hamiltonians
\eqref{1} and \eqref{3} is applicable when the potential of the
superlattice dominates over the Landau quantization:
$\texttt{v}_\perp\texttt{v}_\parallel/l^2_B\ll\Delta^2_\text{eff}$.
This condition is always fulfilled at $l_B\gg d$ ($d=d_\text{I}+d_\text{II}$ is the superlattice period).

In the case of $\Delta_\text{eff}=0$, the found wavefunctions
(positive-frequency solutions) coincide with the wavefunctions
of charge carriers in gapless graphene in a
magnetic field (see \citep{Lozovik} and references therein). In the
case of $\Delta_\text{eff}\neq0$, the zeroth Landau level in each valley
is nondegenerate and all other levels are doubly degenerate.

\begin{center}
3. GREEN'S FUNCTION
\end{center}

To find the Green's function, we will need the
Hamiltonian of a system of noninteracting single-particle
excitations written in terms of secondary quantization
operators:
\begin{equation}\label{13}
\widehat{H}_0=\sum_{n,\,k_y}\widetilde{E}^e_n\widehat{a}^\dagger_{nk_y}\widehat{a}_{nk_y}+\sum_{n,\,k_y}\widetilde{E}^h_n\widehat{b}_{nk_y}\widehat{b}^\dagger_{nk_y},
\end{equation}
where $\widetilde{E}^e_n=E^e_n-V_\text{eff}=\varepsilon_n$ and $\widetilde{E}^h_n=E^h_n-V_\text{eff}=-\varepsilon_n$
(hereinafter, we measure the energy from the level $E=V_\text{eff}$), and $\widehat{a}_{nk_y}$ ($\widehat{a}^\dagger_{nk_y}$) and $\widehat{b}_{nk_y}$ ($\widehat{b}^\dagger_{nk_y}$) are the annihilation
(creation) operators of an electron and hole at the
Landau level with the $y$ component $k_y$ of the crystal
momentum, respectively.

The operators of the electron and hole fields for the
valley of the $K$ point can be written as expansions in
the secondary quantization operators:
\begin{equation}\label{14}
\widehat{\Psi}^{Ke}_{k_y}=\sum_n\Psi^{Ke(+)}_{nk_y}\widehat{a}_{nk_y}+\sum_n\Psi^{Ke(-)}_{nk_y}\widehat{b}^\dagger_{nk_y},
\end{equation}
\begin{equation}\label{15}
\widehat{\Psi}^{Kh}_{k_y}=\sum_n\Psi^{Kh(+)}_{nk_y}\widehat{b}_{nk_y}+\sum_n\Psi^{Kh(-)}_{nk_y}\widehat{a}^\dagger_{nk_y}.
\end{equation}
Here, the arguments $x$ and $y$ are omitted for brevity.
The expressions for the $\Psi$ operators in the valley of the
$K^\prime$ point are similar: they include the wavefunctions
$\Psi^{K^\prime e(\pm)}_{nk_y}(x,\,y)$ and $\Psi^{K^\prime h(\pm)}_{nk_y}(x,\,y)$.

We define the operator of the difference between
the numbers of electrons and holes as
\begin{equation}\label{16}
\widehat{N}_0=\sum_{n,\,k_y}\widehat{a}^\dagger_{nk_y}\widehat{a}_{nk_y}-\sum_{n,\,k_y}\widehat{b}^\dagger_{nk_y}\widehat{b}_{nk_y}.
\end{equation}
Taking into account the commutation relations for the
secondary quantization operators and omitting an
insignificant constant, we write the operator
\begin{equation*}
\widehat{H}^\prime_0=\widehat{H}_0-\widetilde{\mu}\widehat{N}_0
\end{equation*}
\begin{equation*}
=\sum_{n,\,k_y}\left(\varepsilon_n-\widetilde{\mu}\right)\widehat{a}^\dagger_{nk_y}\widehat{a}_{nk_y}+\sum_{n,\,k_y}\left(\varepsilon_n+\widetilde{\mu}\right)\widehat{b}^\dagger_{nk_y}\widehat{b}_{nk_y},
\end{equation*}
where $\widetilde{\mu}=\mu-V_\text{eff}$ and $\mu$ is the chemical potential.

Let us write the $\Psi$ operators in the interaction representation \citep{Tsitovich}
\begin{equation*}
\begin{split}
&\hspace{1.25cm}\widehat{\Psi}^{Ke}_{k_y}(x,\,y,\,t)=e^{i\widehat{H}^\prime_0t}\widehat{\Psi}^{Ke}_{k_y}(x,\,y)e^{-i\widehat{H}^\prime_0t}\\
&=\sum_n\Psi^{Ke(+)}_{nk_y}(x,\,y)\widehat{a}_{nk_y}(t)+\sum_n\Psi^{Ke(-)}_{nk_y}(x,\,y)\widehat{b}^\dagger_{nk_y}(t),
\end{split}
\end{equation*}
\begin{equation*}
\begin{split}
&\hspace{1.25cm}\widehat{\Psi}^{Kh}_{k_y}(x,\,y,\,t)=e^{i\widehat{H}^\prime_0t}\widehat{\Psi}^{Kh}_{k_y}(x,\,y)e^{-i\widehat{H}^\prime_0t}\\
&=\sum_n\Psi^{Kh(+)}_{nk_y}(x,\,y)\widehat{b}_{nk_y}(t)+\sum_n\Psi^{Kh(-)}_{nk_y}(x,\,y)\widehat{a}^\dagger_{nk_y}(t),
\end{split}
\end{equation*}
where
\begin{equation}\label{17}
\begin{split}
\widehat{a}_{nk_y}(t)&=\widehat{a}_{nk_y}e^{-i(\varepsilon_n-\widetilde{\mu})t},\\
\widehat{b}_{nk_y}(t)&=\widehat{b}_{nk_y}e^{-i(\varepsilon_n+\widetilde{\mu})t},
\end{split}
\end{equation}
and $\widehat{a}^\dagger_{nk_y}(t)$ and $\widehat{b}^\dagger_{nk_y}(t)$ are the Hermitian conjugates of
the above two operators.

The Green's function of noninteracting particles is
defined in a standard manner \citep{Gonzalez}:
\begin{equation}\label{18}
\begin{split}
&G^{Ke,h}_{0\alpha\beta}(x,\,x^\prime,\,y-y^\prime,\,t-t^\prime)\\
=-i\langle &T\widehat{\Psi}^{Ke,h}_{k_y\alpha}(x,\,y,\,t)\widehat{\overline{\Psi}}^{Ke,h}_{k_y\beta}(x^\prime,\,y^\prime,\,t^\prime)\rangle,
\end{split}
\end{equation}
where the angle brackets denote statistical averaging; $T$ is the time ordering operator;
$\alpha,\,\beta=1,\,2$ are pseudospin indices; $\widehat{\overline{\Psi}}^{Ke,h}_{k_y\beta}(x^\prime,\,y^\prime,\,t^\prime)=\widehat{\Psi}^{Ke,h\dagger}_{k_y\beta}(x^\prime,\,y^\prime,\,t^\prime)\gamma_0$
is the Dirac conjugation spinor.

Computing averages of the pairs of operators \eqref{17}
and passing from the time $t-t^\prime$ to the frequency $\omega$ and
from the coordinate $y-y^\prime$ to the crystal momentum~$k_y$,
we find the Green's function in the mixed $x-k_y$ representation
\begin{equation}\label{19}
\begin{split}
&\hspace{1.35cm}G^{Ke}_{0\alpha\beta}(x,\,x^\prime;\,k_y,\,\omega)\\
&=\sum_n\frac{\Psi^{Ke(+)}_{nk_y\alpha}(x)\overline{\Psi}^{Ke(+)}_{nk_y\beta}(x^\prime)}{\omega-\varepsilon_n+\widetilde{\mu}-i\delta\sign(\widetilde{\mu}-\varepsilon_n)}\\
&+\sum_n\frac{\Psi^{Ke(-)}_{nk_y\alpha}(x)\overline{\Psi}^{Ke(-)}_{nk_y\beta}(x^\prime)}{\omega+\varepsilon_n+\widetilde{\mu}-i\delta\sign(\widetilde{\mu}+\varepsilon_n)}.
\end{split}
\end{equation}
Hereinafter, we take for brevity the wavefunctions
without the factors $e^{\pm ik_yy}/\sqrt{L_y}$\,; $\delta\rightarrow+0$.

It should be mentioned that Green's function \eqref{19}
formally coincides with the Green's function of a relativistic
electron found with the use of $\Psi$ operators in
the \textit{Furry representation} if one sets $\widetilde{\mu}=0$ \citep{Berestetskii}.

The expression for the Green's function of a hole is
similar to Eq. \eqref{19}: the numerators of the fractions
include the respective products of the hole wavefunctions
and $\widetilde{\mu}$ stands in the denominator with the opposite
sign. Taking into account relation \eqref{10} between the
solutions for different valleys, we find the Green's
functions of an electron and a hole in the valley of the
$K^\prime$ point
\begin{equation}\label{20}
G^{K^\prime e,h}_0(x,\,x^\prime;\,k_y,\,\omega)=-\sigma_xG^{Ke,h}_{0}(x,\,x^\prime;\,k_y,\,\omega)\sigma_x.
\end{equation}

\begin{center}
4. POLARIZATION OPERATOR
\end{center}

The polarization operator is given by the loop diagram
(\hyperlink{fig2}{Fig. 2}). In the case of electrons, it reads \citep{Ratnikov2}
\begin{equation}\label{21}
\begin{split}
&\hspace{0.9cm}\Pi^e(x,\,x^\prime;\,k_y,\,\omega)=-igd\int\frac{dp_y}{2\pi}\int\frac{d\varepsilon}{2\pi}\\
\hspace{-0.25cm}\times\Tr&\left\{\gamma_0G^{Ke}_0(x,\,x^\prime;\,p_y,\,\varepsilon)\gamma_0G^{Ke}_0(x,\,x^\prime;\,p_y+k_y,\,\varepsilon+\omega)\right\},
\end{split}
\end{equation}
where $g=g_sg_v$ is the degeneracy ($g_s=2$ is the spin
degeneracy and $g_v=2$ is the valley degeneracy).

Here, for definiteness, we wrote the electron
Green's functions in the valley of the $K$ point. It is easily
verified using relation \eqref{20} that Eq. \eqref{21} coincides
with the expression for $\Pi^e(x,\,x^\prime;\,k_y,\,\omega)$ written in terms
of the electron Green's functions in the valley of the $K^\prime$
point.

The polarization operator $\Pi^h(x,\,x^\prime;\,k_y,\,\omega)$ for holes
can be written in terms of the hole Green’s function
similar to Eq. \eqref{21}. It is easily seen from the form of the
hole Green’s function and relation \eqref{11} between the
hole and electron solutions that $\Pi^h(x,\,x^\prime;\,k_y,\,\omega)$ differs
from $\Pi^e(x,\,x^\prime;\,k_y,\,\omega)$ only in the sign in front~of~$\widetilde{\mu}$. The~polarization operator of holes is also identical for both
valleys.

As in the case of zero magnetic field \citep{Ratnikov2}, polarization
operator \eqref{21} must be renormalized, since it does
not vanish in the absence of charge carriers, when $|\widetilde{\mu}|~<~\Delta_\text{eff}$. We impose the renormalization condition
\begin{equation}\label{22}
\begin{split}
&\hspace{0.2cm}\Pi^e_\text{Ren}(x,\,x^\prime;\,k_y,\,\omega)\\
=\Pi^e(x,\,x^\prime;&\,k_y,\,\omega)-\left.\Pi^e(x,\,x^\prime;\,k_y,\,\omega)\right|_{|\widetilde{\mu}|<\Delta_\text{eff}}.
\end{split}
\end{equation}
The polarization operator has the form
\begin{equation}\label{23}
\Pi^e_\text{Ren}(x,\,x^\prime;\,k_y,\,\omega)=gd\int\frac{dp_y}{2\pi}F(\xi,\,\xi^\prime;\,\eta,\,\eta^\prime),
\end{equation}
where
\begin{equation*}
\begin{split}
\xi&=\sqrt{\frac{\texttt{v}_\parallel}{\texttt{v}_\perp}}\left(\frac{x}{l_B}+l_Bp_y\right),\\
\xi^\prime&=\sqrt{\frac{\texttt{v}_\parallel}{\texttt{v}_\perp}}\left(\frac{x^\prime}{l_B}+l_Bp_y\right),
\end{split}
\end{equation*}
and the variables $\eta$ and $\eta^\prime$ differ from $\xi$ and $\xi^\prime$, respectively,
in the replacement $p_y\rightarrow p_y+k_y$.

\begin{figure}[t!]
\begin{center}
\hypertarget{fig2}{}
\includegraphics[width=0.46\textwidth]{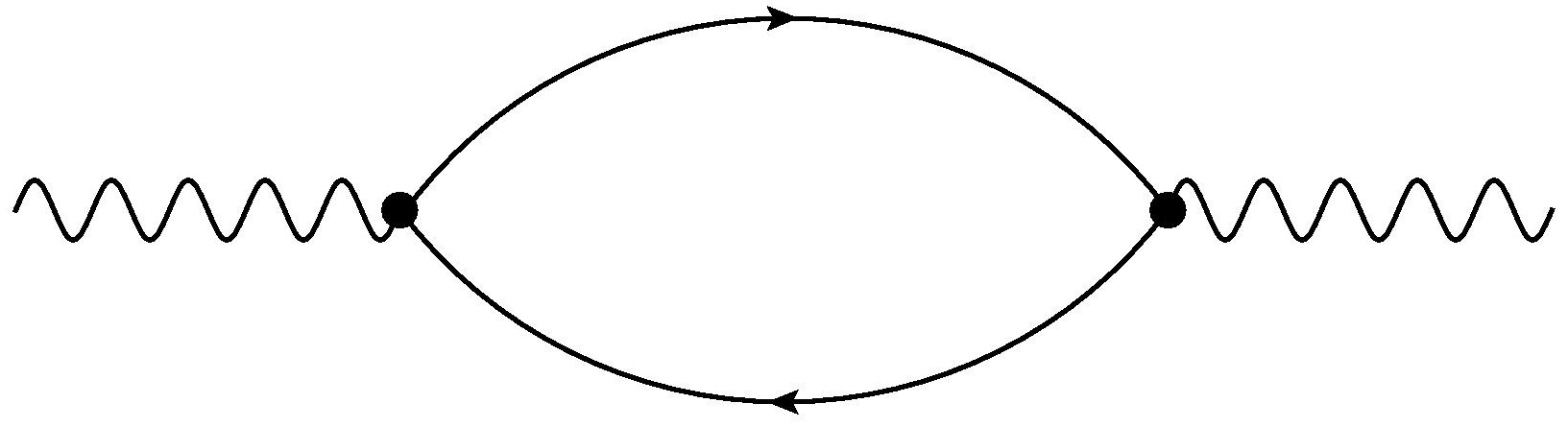}
\end{center}

\begin{list}{}
{\rightmargin=0.27cm\leftmargin=0.27cm}
\item
\footnotesize{\bf Fig. 2.} Loop diagram. The vertices are the matrices $\gamma_0$.
\end{list}
\normalsize
\end{figure}

The function $F(\xi,\,\xi^\prime;\,\eta,\,\eta^\prime)$ is the simplest if only
the zeroth Landau level is filled:
\begin{equation}\label{24}
\begin{split}
&\hspace{2.28cm}F(\xi,\,\xi^\prime;\,\eta,\,\eta^\prime)\\
&=\sum_{s=\pm}\sum_{n=1}^\infty\frac{2(\varepsilon_n-s\varepsilon_0)}{\omega^2-(\varepsilon_n-s\varepsilon_0)^2+si\delta\sign\omega}\\
&\hspace{0.65cm}\times|C^{(s)}_n|^2\phi_n(\xi)\phi_n(\xi^\prime)\phi_0(\eta)\phi_0(\eta^\prime).
\end{split}
\end{equation}
Here, the sum at $s=+$ is the contribution of positive-frequency
solutions (virtual intra-miniband transitions),
whereas the sum at $s=-$ is the contribution of
negative-frequency solutions (virtual inter-miniband
transitions).

There is an additional significant simplification:
the integrals of $\phi_n(\xi)\phi_n(\xi^\prime)\phi_0(\eta)\phi_0(\eta^\prime)$ over $p_y$ depend
only on $x-x^\prime$, as for charge carriers with a quadratic
dispersion relation \citep{Lerner}. Calculating these integrals
and performing the Fourier transform from $x-x^\prime$ to
$k_x$, we find the polarization operator in the momentum
representation
\begin{equation}\label{25}
\begin{split}
&\hspace{1.65cm}\Pi^e_\text{Ren}({\bf k},\,\omega)=\frac{gd}{2\pi l^2_B}\\
&\times\sum_{s=\pm}\sum_{n=1}^\infty\frac{2(\varepsilon_n-s\varepsilon_0)}{\omega^2-(\varepsilon_n-s\varepsilon_0)^2+si\delta\sign\omega}\\
&\hspace{1.3cm}\times\frac{|C^{(s)}_n|^2}{n!}\left(\frac{\chi^2}{2}\right)^ne^{-\chi^2/2},
\end{split}
\end{equation}
where
\begin{equation*}
\chi^2=\frac{\texttt{v}^2_\perp k^2_x+\texttt{v}^2_\parallel k^2_y}{\texttt{v}_\perp\texttt{v}_\parallel}l^2_B.
\end{equation*}

\begin{center}
5. DISPERSION RELATION\\
OF MAGNETOPLASMONS
\end{center}

For simplicity, we restrict ourselves to the case of
occupation of only the zeroth Landau level (of electrons
or holes). The dispersion relation of collective
excitations in plasma in the random-phase approximation
is determined by the equation

\begin{equation}\label{26}
1-V({\bf k})\Pi^e_\text{Ren}({\bf k},\,\omega)=0,
\end{equation}
where $V({\bf k})$ is the Coulomb interaction between the
charge carriers in the superlattice.

In our case, it is the same as for semiconductor filaments
arranged periodically in the same plane parallel
to each other. The Coulomb interaction between charges
in two filaments separated by the distance $\nu d$
in such a system is \citep{Andryushin}
\begin{equation}\label{27}
V(\nu,\,k_y)=2\widetilde{e}^2K_0\left(d|\nu k_y|\right),
\end{equation}
where $d$ is the distance between the gapless graphene
strips (coincides with the superlattice period); $\nu$ is the
strip (superlattice cell) number; $\widetilde{e}^2=e^2/\kappa_\text{eff}$, where
$\kappa_\text{eff}=(\kappa_1+\kappa_2)/2$ is the effective static dielectric constant
determined by the static dielectric constants $\kappa_1$
and $\kappa_2$ of the media surrounding graphene, for example,
vacuum and the substrate material; and $K_0(x)$ is
the modified Bessel function of the second kind.

Let us pass from the discrete variable of the strip
number $\nu$ to the transverse momentum $k_x$ ($-\pi/d\leq k_x\leq\pi/d$)
similar to \citep{Andryushin}
\begin{equation}\label{28}
\begin{split}
&\hspace{2cm}V({\bf k})=\sum_{\nu=-\infty}^\infty V(\nu,\,k_y)e^{i\nu k_xd}\\
&=2\widetilde{e}^2K_0\left(\frac{d_\text{I}}{2}|k_y|\right)+4\widetilde{e}^2\sum_{\nu=1}^\infty\cos(\nu k_xd)K_0\left(\nu d|k_y|\right),
\end{split}
\end{equation}
where $d_I$ is the width of the gapless graphene strips.

Expression \eqref{28} is simplified in the case of interest
of a small gapped graphene strip width $d_\text{II}\ll d_\text{I}$ \citep{Andryushin}
\begin{equation}\label{29}
\begin{split}
&\hspace{2.55cm}V({\bf k})=2\widetilde{e}^2\ln\frac{d}{\pi d_\text{I}}\\
&+\left[-2C-2\psi\left(\frac{k_xd}{2\pi}+\frac{1}{2}\right)+\pi\tan\frac{k_xd}{2}\right]\widetilde{e}^2+o(k_yd),
\end{split}
\end{equation}
where $C=0.577\ldots$ is the Euler constant and $\psi$ is the
Euler $\psi$ function. At the edges of the miniband ($k_x=\pm\pi/d$), we find from Eq. \eqref{28}
similar to \citep{Andryushin}
\begin{equation}\label{30}
V({\bf k})=2\widetilde{e}^2\ln\frac{d}{\pi d_\text{I}}+\frac{2\pi\widetilde{e}^2}{|k_y|d}+o(k_yd).
\end{equation}

\begin{center}
6. NUMERICAL CALCULATION\\
OF MAGNETOPLASMON FREQUENCIES
\end{center}

For numerical calculations, we take as an example
a graphene–graphane superlattice formed by alternating
strips of gapless graphene with the width $d_\text{I}=8.52$~nm
and graphane strips with the width $d_\text{II}=0.852$~nm.
The superlattice period is $d=d_\text{I}+d_\text{II}=9.372$~nm.

For graphane, $\Delta=2.7$ eV \citep{Lebegue}. For simplicity, we
set $V_\text{eff}=0$. Then we find from the calculation of the
dispersion relation of charge carriers $\Delta_\text{eff}=98.56$~meV,
whereas $E=102.76$~meV at the edge of the lower electron
miniband. Let the chemical potential be $\widetilde{\mu}=100$~meV. Then, $\texttt{v}_\perp\approx0.19\times10^8$ cm/s and $\texttt{v}_\parallel\approx\texttt{v}_\text{F}=0.85\times10^8$ cm/s near $E=\widetilde{\mu}$ \citep{Elias}. The magnetic field is
$B=5$ T. The magnetic length amounts to $l_B=11.47$~nm.
Only the zeroth Landau level is occupied: $\varepsilon_0=\Delta_\text{eff}$, $\varepsilon_1=103.79$~meV and $\varepsilon_0<\widetilde{\mu}<\varepsilon_1$. The effective
static dielectric constant is $\kappa_\text{eff}=5$.

\begin{figure}[t!]
\begin{center}
\hypertarget{fig3}{}
\includegraphics[width=0.5\textwidth]{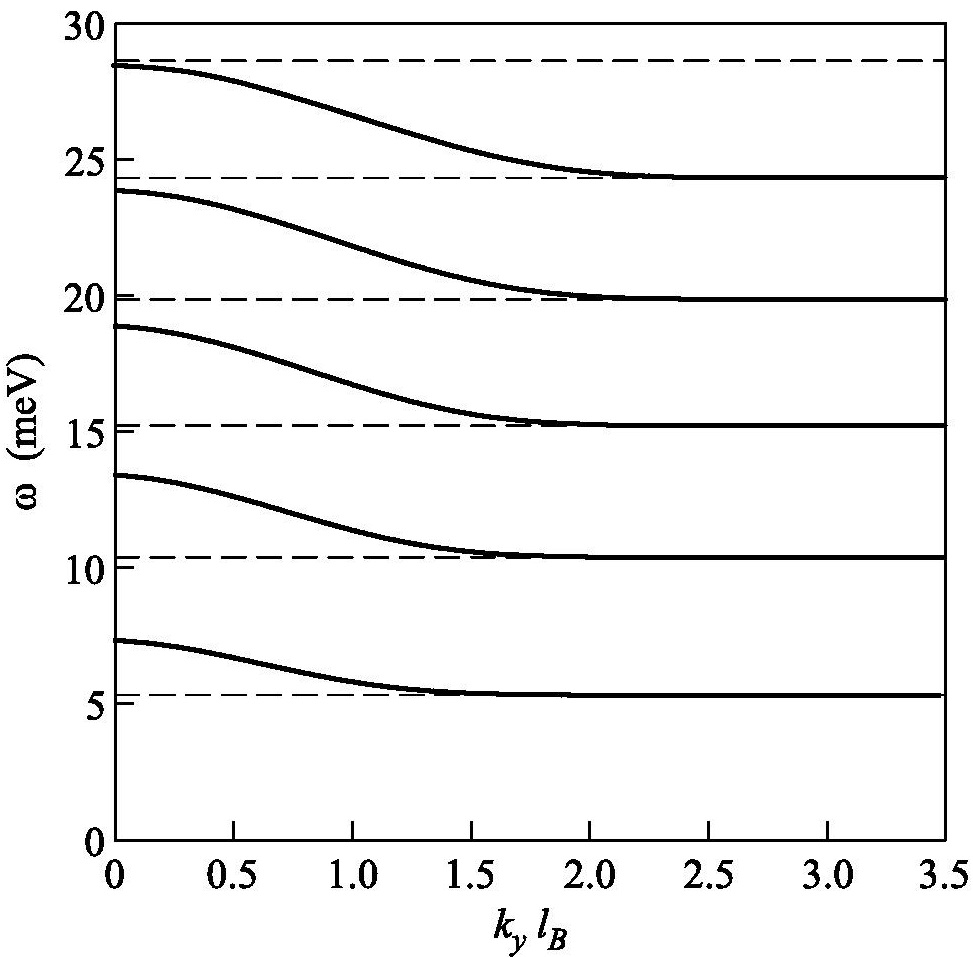}
\end{center}

\begin{list}{}
{\rightmargin=0.27cm\leftmargin=0.27cm}
\item
\footnotesize{\bf Fig. 3.} Dispersion of magnetoplasmons in a graphene–gra-phane superlattice at the edge of a miniband.
\end{list}
\normalsize
\end{figure}

Let us calculate the dependence of the magnetoplasmon
frequencies at the edge of the miniband, $k_x=\pm\pi/d$,
on the crystal-momentum component $k_y$. The
results for five lower branches of the magnetoplasmon
spectrum are shown in \hyperlink{fig3}{Fig. 3}. Horizontal dashed~lines
mark the resonance frequencies corresponding to a
non-zero imaginary part of the polarization operator:
$\omega_n=\varepsilon_n-\varepsilon_0$ for $n=1,\,2,\,\ldots,\,6$. The dispersion curves of
magnetoplasmons do not intersect these horizontal lines, just tending to them asymptotically at $k_yl_B\gg1$.

\begin{center}
7. CONCLUSIONS
\end{center}

In this work, the problem of dispersion of magnetoplasmons
in planar graphene-based superlattices has
been studied analytically in the random-phase
approximation. Since the lower electron miniband
and the upper hole miniband are situated pretty close
to one another in energy, the superlattice behaves like
an anisotropic narrowband semiconductor. The
Green's function of charge carriers has been derived in
a standard way. The polarization operator has been
calculated from the found Green's function in the
zeroth approximation with respect to the interaction.
Apart from the contribution of virtual intra-miniband
transitions, the polarization operator includes the
contribution of virtual inter-miniband transitions,
which allows taking into account this contribution
explicitly in the dispersion relation of magnetoplasmons
in the medium under consideration.

I am grateful to A.P. Silin for the discussion and
valuable advice regarding this publication.

\begin{flushright}
\emph{Translated by A. Safonov}
\end{flushright}
\end{document}